\documentclass[a4paper,fleqn,usenatbib]{mnras}

\usepackage{mathptmx}

\usepackage[T1]{fontenc}
\usepackage{ae,aecompl}

\usepackage{graphicx}	
\usepackage{amsmath}	
\usepackage{amssymb}	

\title[An explanation of the periphery of Tycho's SNR]{An explanation of the formation of the peculiar periphery of Tycho's supernova remnant}

\author[J. Fang et al.]{
Jun Fang,$^{1}$\thanks{E-mail: fangjun@ynu.edu.cn (JF)}
Huan Yu,$^{2}$\thanks{E-mail: yuhuan.0723@163.com (HY)}
and Li Zhang$^{1}$\thanks{E-mail: lizhang@ynu.edu.cn (LZ)}
\\
$^{1}$Department of Astronomy, Key Laboratory of Astroparticle Physics of Yunnan Province, Yunnan University, Kunming 650091, China\\
$^{2}$Department of Physical Science and Technology, Kunming University, Kunming 650214, China
}

\date{Accepted XXX. Received YYY; in original form ZZZ}

\pubyear{2017}

\begin{document}
\label{firstpage}
\pagerange{\pageref{firstpage}--\pageref{lastpage}}
\maketitle

\begin{abstract}
Tycho's supernova remnant (SNR) has a periphery which clearly deviates from spherical shape bash on the X-ray and radio observations. The forward shock from the southeast to the north of the remnant has a relatively deformed outline with a depression in the east, although in the west it is generally round and smooth. Moreover, at some locations of the shell, the supernova ejecta distributes closely to the forward shock to produce protrusions.  Using 3D hydrodynamical simulation, the dynamical evolution of the supernova ejecta in an inhomogeneous medium and the formation process of the profile of the forward shock can be studied. To reproduce the peculiar periphery of the remnant, we propose a model in which the supernova ejecta has evolved in a cavity blown by a latitude-dependent outflow. The results indicate that, with the assumptions of the wind bubble driven by the anisotropic wind with a mass loss rate of $\sim10^{-7} {\rm M_{\odot}}$~yr$^{-1}$, a pole velocity of $\sim 100$~km s$^{-1}$, a duration of $\sim10^5$~yr just before the supernova explosion, and a spatial velocity of $\sim 30$~km s$^{-1}$ of the progenitor with respect to the circumstellar medium, the depression to the east and the  protrusion to the southeast on the observed periphery of the remnant can be generally reproduced. In conclusion, an explanation to the peculiar shape of the periphery of Tycho's SNR is the supernova ejecta evolved in the cavity driven by the latitude-dependent wind.
\end{abstract}




\section{Introduction}
\label{intro}
Tycho's SNR (G120.1+1.4) stems from a type Ia supernova (SN) occurred in AD 1572 \citep{SG02} based on the spectroscopic analyses of X-rays \citep{Bea06} and light
echo \citep{Rea08,Kea08}, and the distance reported in the literatures ranges from $2.3\,\mathrm{kpc}$ based on optical observations \citep{Cea80}, $2.5\,\mathrm{kpc}$ derived from HI absorption
studies \citep{TL11} to $4.0\,\mathrm{kpc}$ based on higher-resolution HI data \citep{Sea95}. It has a generally round shell-type morphology indicated
at radio \citep{Dea82}, optical \citep{Gea00} and X-ray wavelengths \citep{Hea02}. Moreover, it is also a GeV/TeV $\gamma$-ray source based on the observations with {\it Fermi} \citep{Gea12} and VERITAS \citep{Aea11}.  As indicated in either the {\it Chandra} X-ray or the VLA radio image, although the periphery is generally
regular, two bulges which are separated by a depression are clear located at the northeast and the southeast of the shell, respectively  \citep{Eea11,Wea16}. Furthermore, the proper motion of the forward shock varies with
direction, and the expansion velocities of the western and the southwestern edges of the forward shock are higher than the east and the northeast ones by
a factor of two \citep{Kea10,Wea16}. The reason for the peculiar properties of the remnant remains an open question.

Aside from the morphology of the remnant, it is still in debate whether the progenitor of Tycho's SNR is single-degenerate (SD) or double-degenerate (DD). In the SD scenario, a white dwarf explodes when the mass reaches the Chandrasekhar limit due to accreting matter from a companion star \citep{WI73,WH12,Wea14}. Alternately, two white dwarfs merge to trigger the supernova (SN) in the DD case \citep{W84}. The SD model for the progenitor of Tycho's SNR is favored with evidences including the discoveries of both a candidate (Tycho-G) of the companion star \citep{R04} and an X-ray arc inside the remnant, which is theoretically originated from the interaction between the ejecta and the companion star's envelope \citep{Lea11}. Furthermore, an expanding molecular bubble has been indicated to be associated with Tycho's remnant based on the CO observations \citep{Zea16,Cea16}. The bubble is argued to be the result of the outflow driven by the accretion between the white dwarf and the companion star, and the progenitor model of SD is supported in this scenario.  However, the argument that the star Tycho-G is the donor star was challenged based on both the detected proper motion of Tycho-G \citep{Kea08} and a significant offset of the explosion site from the geometric center based on the azimuthal variations of the proper motion of the periphery of the remnant \citep{Rea97,H00,Kea10,XS15,Wea13,Wea16}. Recently, \citet{Woea17} reported stringent constraints on hot, luminous progenitors for Tycho's SNR, and the DD scenario was favored due to the lack of a Str\"{o}mgren sphere.

Based on MHD/HD simulations, the SNR morphology can be used to trace the matter distribution of the circumstellar medium, which can shed light on the progenitor of the remnant. For example, \citet{Tea14} assumed Kepler's SNR had been evolved inside the wind bubble which was produced as the interaction between the wind from an AGB companion star and the interstellar matter based on 3D HD simulation, and the resulting morphology was consistent with that indicated in X-rays. Another SNR G1.9$+$0.3 was argued to be a SN Ia exploded in an elliptical planetary nebula to explain the X-ray morphology \citep{TS15}. \citet{FYZ17} numerically studied the morphology of Cygnus Loop whose X-ray morphology indicated irregular features, such as a blowout in the south, and a bump in the west. The peculiar morphology of the remnant can be generally reproduced with the assumption that it had evolved  in a cavity produced by the anisotropic wind. There was an evidence for a warm ISM  cloud located in the east of Tycho's SNR \citep{Gea00}, and the matter around it is nonuniform with azimuthally-varied density \citep{Wea13}.  For the remnant, 2D HD simulations had been performed with a nonuniform medium with a density gradient to investigate the offset of the explosion site from the geometric center \citep{Wea13,Wea16}.

In this paper, the evolutions of both the wind bubble and the supernova ejecta in the bubble for Tycho's SNR are investigated using 3D HD simulations to explain the observed periphery of the remnant. In the model, just before the SN explosion, an anisotropic wind with a latitude-dependent density, which can be interpreted as the wind due to the accretion, is blown into the ambient matter, and the system has a spatial velocity to the west. As a consequence, similar as the observed image as indicated in radio and X-rays, a profile of the forward shock which has a generally round shape in the west, a depression in the east, and a prominent protrusion from the east to southeast is reproduced.
This paper is organized as follows. In Section \ref{model}, the numerical model is presented. The simulation results are given in Section \ref{sim:results}. Finally, the main conclusions and some discussion are shown in Section \ref{DISCON}.

\section{The Model and Numerical Setup}
In this section, the model for the formation of the wind-blown cavity is reviewed, and the numerical setup of the supernova ejecta is also presented.
\label{model}
\subsection{Modeling the wind from the progenitor system and the supernova ejecta}
\begin{figure}
\begin{center}
\includegraphics[width=0.4\textwidth]{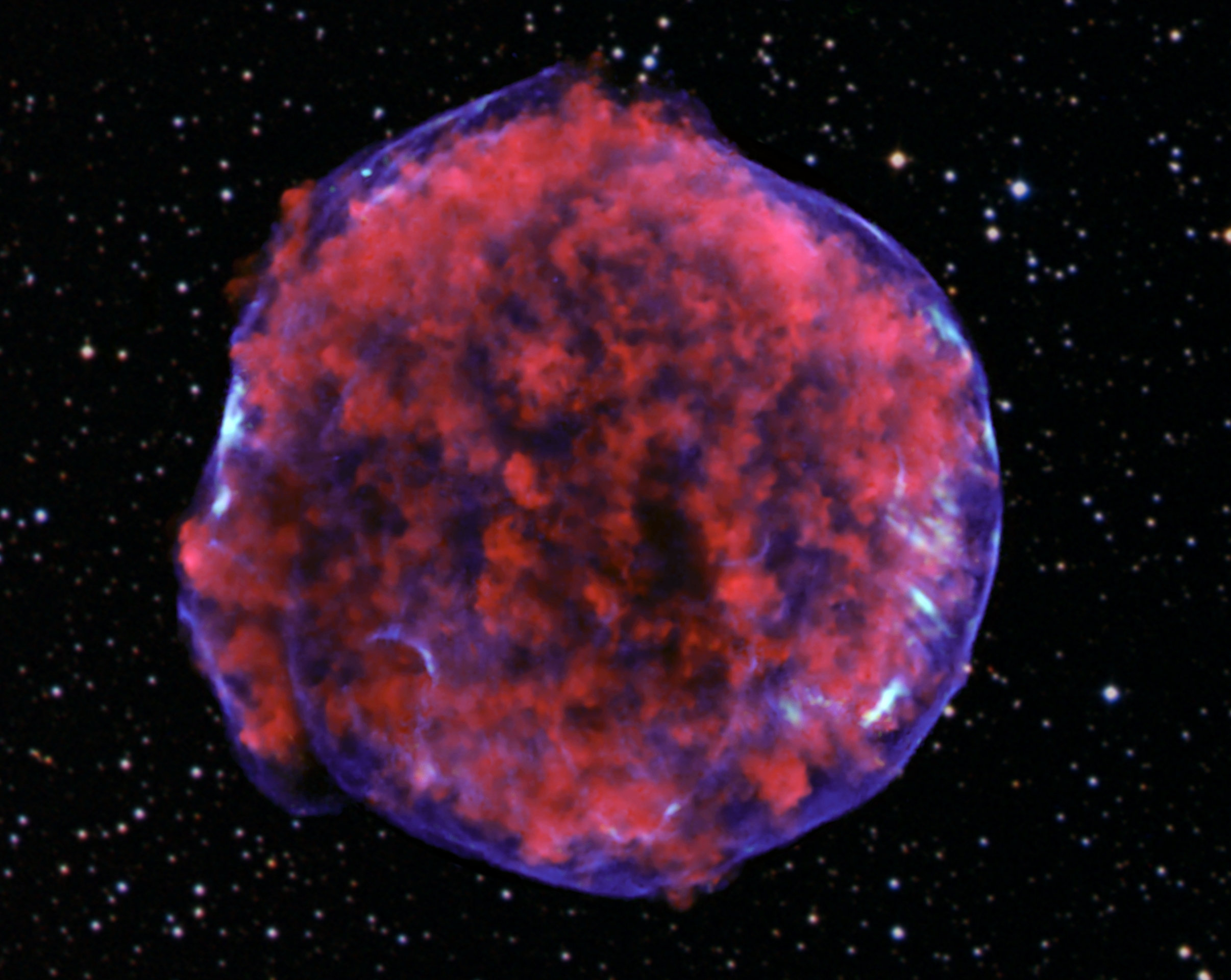}
\caption{The observed X-ray morphology of Tycho's SNR with Chandra including low-energy (red) and high-energy (blue) X-rays \citep{Eea11}. (This image is downloaded from http://chandra.harvard.edu/photo/2011/tycho/) \label{fig:tychoxray}}
\end{center}
\end{figure}

Fig.\ref{fig:tychoxray} shows the X-ray image of Tycho's SNR observed with {\it Chandra} \citep{Eea11}. Lower energy X-rays (red) in the image show expanding debris from the supernova explosion, and high energy X-rays (blue) represent the shock wave where relativistic particles are accelerated by the shock. Tycho's SNR has a deformed shell with a depression in the east, although the west part is generally round. There are a bulge to the north and a prominent protrusion to the southeast of the shell of the remnant. Moreover, the expanding debris distributes closer to the forward shock in the east than in the west. To interpret the peculiar configuration, we assume that a latitude-dependent outflow had been driven from the progenitor system before the SN explosion, and a wind-blown cavity was produced due to the interaction between the wind and the circumstellar matter. Moreover, to reproduce the asymmetry of the remnant, the progenitor is assumed to possess a spatial velocity $v_{\mathrm{s}}$ directing to the west with respect to the local circumstellar matter.

A 3D Cartesian coordinate system is adopted in the simulations with $+z$ directing to the north and $+y$ to the west.  A fast wind can arise due to accretion from a companion star to a WD with a mass loss rate of $\sim 10^{-7} - 10^{-6}\mathrm{M_{\odot}~ yr^{-1}}$ and a wind velocity larger than $100\, \mathrm{km~s^{-1}}$ \citep{KH99,Zea16}. In the simulations, the wind with a mass-loss rate of $\dot{M}_{\mathrm{w}}$ is driven into the ambient uniform medium with a  number density of $n_{\mathrm{0}}$ and a temperature of $T_0$ for a lasting time of $T_{\mathrm{w}}$. To explain the bulge near the north, a depression to the east, and the protrusion to the southeast on the periphery of the remnant, a latitude-dependent distribution of the density, which increases from the equator to the poles with an equator-to-pole ratio $\xi$, for the wind is adopted, i.e.,
\begin{equation}
\rho(r, \theta)=\frac{A}{r^2}f(\theta)\;.
\label{eqs:rho}
\end{equation}
The wind is injected in the circumstellar medium at a spherical border with $r=0.4 \,\mathrm{pc}$, and
\begin{equation}
v_{\rm w}(\theta) = \frac{v_{\rm p}}{\sqrt{f(\theta)}}\;,
\label{eqs:vel}
\end{equation}
$v_{\rm p}$ is the terminal velocity of the wind at the pole,
$\theta$ is the polar angle, and
\begin{equation}
f(\theta)=\xi-(\xi-1)|\cos\theta|^{1/2}\;.
\label{eqs:ftheta}
\end{equation}
The scaling constant $A$ can be calculated with
\begin{equation}
A=\frac{15(\xi-1)^2}{8\xi^{5/2}-20\xi+12}\frac{\dot{M}_{\rm w}}{4\pi v_{\rm p}}\;.
\label{eqs:afactor}
\end{equation}
This configure for the anisotropic wind was originally proposed in \citet{Rea08} to model
the tail of the star Mira with $\xi = 20 $, and it was also employed to simulate the wind-blown cavity around the Kepler's SNR \citep{Tea14} and the Cygnus loop \citep{FYZ17}.

The evolution of the remnant is initiated by setting the ejecta which has a mass of $M_{\mathrm{ej}}=1.4M_{\odot}$, a kinetic energy of $E_{\rm ej}=10^{51}$erg and a radius of $R_{\mathrm{ej}}=0.5$pc in the wind bubble. The inner part of the core of the ejecta has a mass of $\eta M_{\mathrm{ej}}$ with $\eta=4/7$, and the density has a power-law distribution on $r$  with an index of $s=7$ in the outer layer for the remnant of the SN Ia, i.e.,
\begin{equation}
\rho_{\rm ej}(r) = \left\{
  \begin{array}{cc}
    \rho_{\mathrm c}& \mathrm{if~} r < r_{\mathrm c}\;, \\
    \rho_{\mathrm c}(r/r_{\mathrm c})^{-s} & \mathrm{if~} r_{\mathrm c} < r <  R_{\mathrm{ej}}\;,
    \label{rho_c}
   \end{array}
\right.
\end{equation}
and  $r_{\mathrm c} = 0.86R_{\mathrm{ej}}$ according to
\begin{equation}
r_{\mathrm c}=\left[ \frac{3-4\eta}{3(1-\eta)} \right]^{\frac{1}{s-3}} R_{\mathrm{ej}} .
\end{equation}
The other details of the initiation of the ejecta can be seen in \citet{JN96}.

\subsection{Numerical method}
The dynamical evolutions of both the wind-blown cavity and the SNR are performed using 3D HD simulations with the PLUTO code \citep{Mea07,Mea12} in a 3D Cartesian coordinate system. The dynamical properties are investigated by solving the Euler equations of gas dynamics, i.e.,
\begin{eqnarray}
\frac{\partial\rho}{\partial t} + \nabla\cdot(\rho \textbf{v}) & = & 0\; , \\
\frac{\partial \rho {\bf v}}{\partial t} + \nabla \cdot \rho {\bf
    vv}   + \nabla{P} & = & 0\; , \\
\frac{\partial E}{\partial t} + \nabla \cdot (E+P){\bf v} ) & = & - \left ( \frac{\rho}{m_{\rm H}}\right)^2 \Lambda(T)\; ,
\end{eqnarray}
where $P $ is the gas pressure and $E$ is the total energy density
\begin{equation}
E = \frac{P}{\gamma - 1}+\frac{1}{2}\rho v^2 \;,
\nonumber
\end{equation}
$t$ is the time, $\rho = \mu m_{\rm H} n$ is the mass density, the molecular weight $\mu$ is $0.6$ for the ionized gas with an assumption
of a 10:1 H:He ratio,
$m_{\rm H}$ is  the mass of the hydrogen atom,  the adiabatic index $\gamma$ is adopted to be $5/3$
for the nonrelativistic gas, $\bf{v}$ is the gas velocity, and $\Lambda(T)$ is the
tabulated cooling function \citep{Mea07,Mea12}. 

The simulations are performed with the computational domain of $12\times12\times12$~pc with $512^3$ grids. The observations with $\it Spitzer$ indicated significant variation in density around the periphery of Tycho's SNR, and a mean preshock density of $0.08 - 0.1\,\mathrm{cm}^{-3}$ was derived when the three regions associated with dense knots were ignored \citep{Wea13}. In this paper, $n_0 = 0.1$cm$^{-3}$ is adopted as the density of the ambient  proton, and we assume that the wind is continuously blown into the ambient medium with a temperature of $T_0=10^4$\,K. The wind has been injected for a time of $T_{\mathrm{w}}$, and then a distorted wind bubble is generated as a result of the interaction between the circumstellar medium and the wind from the progenitor system which has a spatial velocity of $v_{\mathrm{s}}$ to the $+y$ direction.  Then the ejecta is imposed in the center of the computational domain, and the Euler equations are solved for $T_{\mathrm{age}}=450$~yr, which is the time after the injection of the ejecta into the wind bubble, to simulate the dynamical evolution of the remnant in the bubble.

\section{Results}
\label{sim:results}

\begin{table}
 \renewcommand{\arraystretch}{1.3}
  \caption{Parameters for the numerical models. The common parameters are $n_{\rm 0}$ = $ 0.1 \rm cm^{-3}$, $T=10^4$ K, $v_{\rm p}=100\rm km\,s^{-1}$,
   $\dot{M}_{\rm w} = 10^{-7} {\rm M}_{\odot} {\rm yr}^{-1}$,  $\xi=20$,
   $M_{\rm ej}=1.4 {\rm M_{\odot}}$,
    and $E_{\rm ej} = 10^{51}$ erg.}
    \centering
  \label{table:model}
  \begin{tabular}{ccc}
    \hline\hline
  Parameter   & $\alpha$  ($^{\circ}$)   & $v_{\rm s}$ ($\rm km\,s^{-1}$)     \\
    \hline
    Model A   & 40                       & 30     \\
    Model B   & 50                       & 30      \\
    Model C  & 60                       & 30      \\
   Model D   & 50                       & 20      \\
    \hline
  \end{tabular}
\end{table}

\subsection{Evolution of the wind}

\begin{figure*}
\begin{center}
\includegraphics[width=1.0\textwidth]{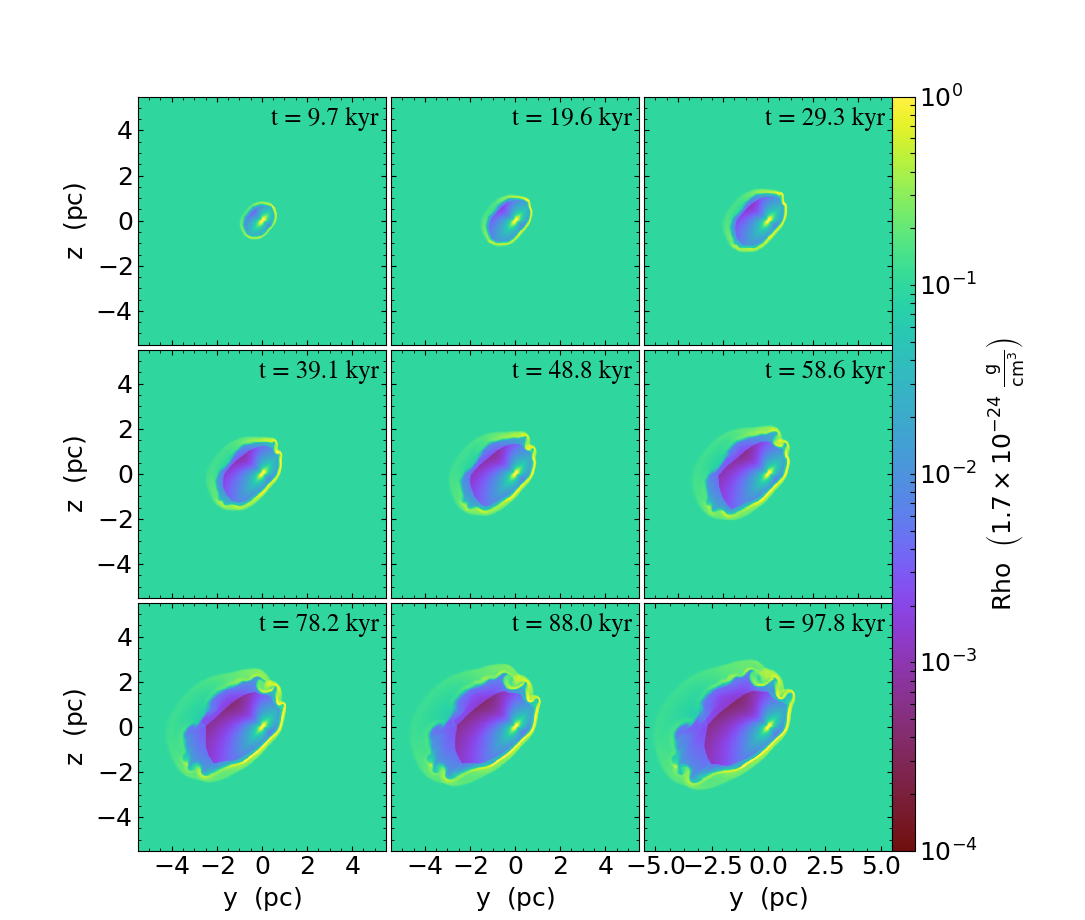}
\caption{Evolution of the wind bubble for Model B. The $yz$ slices of the mass density in units of $m_\mathrm{H}~\mathrm{cm^{-3}}$ at $x=0$ for different integration times. The time in the panels is measured from the start of the wind, up to the explosion time $t=T_{\rm w}$. \label{fig:windevo}}
\end{center}
\end{figure*}

Assuming that the wind originates from the outflow driven by the accretion with the parameters $v_{\rm p}=100~\rm \mathrm{km\,s}^{-1}$,
$\dot{M}_{\rm w} = 1.0\times10^{-7} {\rm M}_{\odot}~\mathrm{yr}^{-1}$, it is continually injected at a radius of $0.4~\mathrm{pc}$ into the ambient uniform medium with $n_0=0.1~\mathrm{cm}^{-3}$ and $T_0=10^4~\mathrm{K}$. Moreover, the ambient medium has a velocity of $v_{\mathrm{s}}=30~\mathrm{km\,s}^{-1}$ (Model B in Table \ref{table:model}) towards the $-y$ direction to simulate the progenitor system moving to the $+y$ direction. Fig.\ref{fig:windevo} shows the evolution of the wind which is latitude-dependent with $\xi=20$ and the angle from the polar direction of the star to the $+y$ direction $\alpha=50^{\circ}$ for an integral time up to $9.78\times10^4~\mathrm{yr}$. A bubble surrounded by a shock with a distorted oval shape is generated as the wind propagating into the ambient matter. The shock has an elongated and distorted morphology as a joint result of the anisotropic wind and the motion of the progenitor relative to the circumstellar medium.

During the evolution of the wind bubble, aside from the outer shock with a disturbed oval shape, a terminate shock around the stellar wind is produced. The outer shock and the terminate shock compress the ambient material and the stellar wind, respectively. They are close to each other in the west, but the separation becomes larger towards the east. More material is injected in the equator of the wind, whereas the poles are relatively tenuous. As a result, the bubble is elongated along the equator, and a deformed oval morphology is produced.  The stagnation distance estimated by balancing momentum fluxes (see Eq.(1) in \citet[]{Tea14}) is $0.6$ pc, and it is clearly illustrated in Fig.\ref{fig:windevo} that the outer shock in the west can extend to this distance. Oppositely, it is more extended in the $-y$ direction due to the spatial velocity of the system relative to the ISM. After a time of $\sim 3\times10^4~\mathrm{yr}$, Kelvin-Helmholtz instabilities arise significantly  on the border of the shock due to density and velocity differences between the wind and the ambient matter. Several protrusions on the northwestern border of the shock are produced, and they can influence the shape of the forward shock of the SNR evolving inside the bubble.

\subsection{Evolution of the SNR}

\begin{figure*}
\begin{center}
\includegraphics[width=0.8\textwidth]{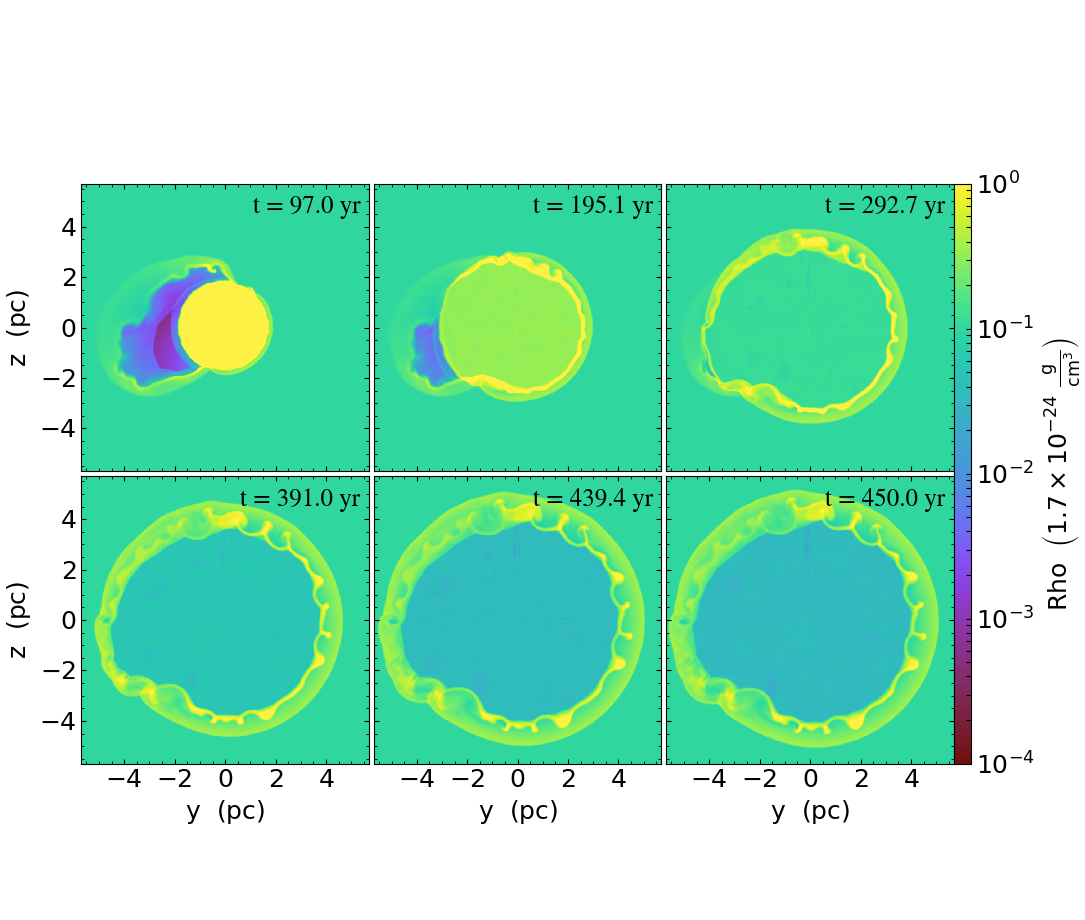}
\caption{Evolution of the SNR in the wind bubble for Model B. The $yz$ slices of the mass density in units of $m_\mathrm{H}~\mathrm{cm^{-3}}$ at $x=0$ for different integration times after the supernova explosion are indicated. $\rm t$ in the panels is the time after the injection of the ejecta, up to the present time of $T_{\rm age}$.
 \label{fig:snrevo}}
\end{center}
\end{figure*}

\begin{figure}
\begin{center}
\includegraphics[width=0.5\textwidth]{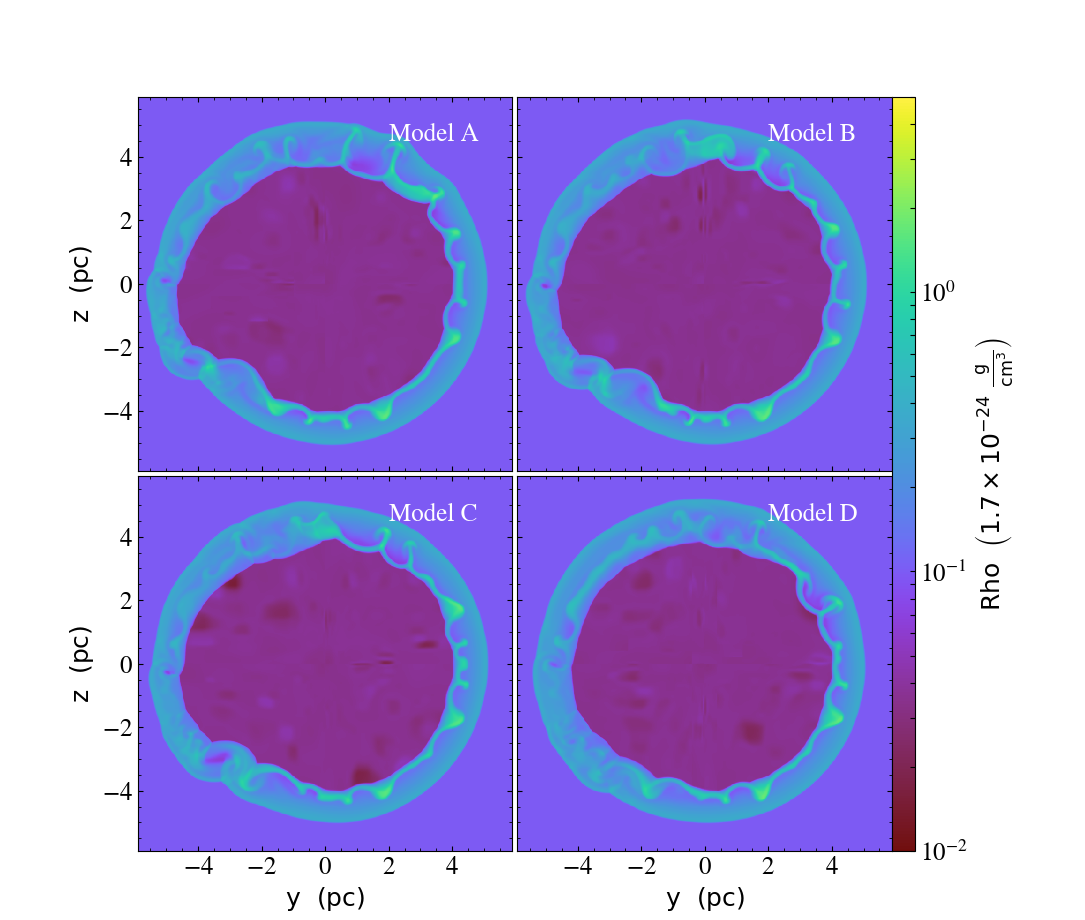}
\caption{The resulting density in units of $m_\mathrm{H}~\mathrm{cm^{-3}}$ of the SNR in the plane $x=0$ for the Model A, B, C, and D for $\mathrm{T_{age}} = 450\,\mathrm{yr}$, respectively. \label{fig:modelabcd}}
\end{center}
\end{figure}

In model B, after the anisotropic wind has been injected for a time of $T_{\mathrm{w}}=9.78\times10^4~\mathrm{yr}$, the supernova ejecta is set at the center of the simulation domain to simulate the evolution of the remnant in the wind bubble.
A forward shock and a reverse shock are produced when the  high-density and high-velocity ejecta transports into the wind bubble, and Rayleigh-Taylor instabilities are developed near the location of the contact discontinuity between the two shocks.
Because the ejecta has a significant displacement from the center of the bubble due to the motion of the progenitor relative to the ambient medium, the forward shock expanding into the west ($+y$ direction) initially overtakes the terminate shock and then it transports into the ambient uniform medium at a time smaller than $100~\mathrm{yr}$ in this direction. Alternatively, the forward shock arrives at the border of the wind bubble at a later time in the east. Especially, in the southeast, the forward shock of the remnant evolves in the tenuous wind bubble during a longer time of $\sim 250~\mathrm{yr}$. At an age of $\sim450~\mathrm{yr}$, the western part of the forward shock has a generally round morphology although the eastern shell is distorted with one depression in the east and a prominent protrusion in the southeast.

For the deformed eastern shell, and a prominent bump appears on the southeastern part of the shell in the figure  at a time of $\sim 450~\mathrm{yr}$. Moreover, the profile from the northeast to the north of the shell is relatively regular, and a minor bulge appears near the north pole of the shell.  These properties are consistent with the observed morphology of Tycho's SNR in either the X-ray or the radio band.
With $\alpha=50^{\circ}$, the wind bubble is elongated along the direction from southeast to northwest. As the SNR shock propagates in the southeast, it encounters the dilute wind material in a longer journey compared with the other directions, and then a bump is produced in this direction. This bump, which spans from the east depression to the southeastern edge which is coincident with the south pole of the wind bubble, has an angular extension of about $40^{\circ}$.

At the forward shock, nonthermal X-rays are effectively produced because electrons have been accelerated to relativistic energies by the shock, and it can be concluded from the detected X-ray morphology that the distance of the forward shock to the thermal plasma has a tendency of decreasing from west to east. Especially, the shocked ejecta distributes very closely to the forward shock in the southeastern bump, and this tendency is also reproduced from the simulation.

The influences of the inclination angle $\alpha$ and the spatial velocity $v_{\mathrm s}$ on the shape of the remnant are illustrated in Fig.\ref{fig:modelabcd} for the Models A, B, C, D at $t = 450 \, \mathrm{yr}$, respectively, and the parameters for the different models are  shown in Table \ref{table:model}. With a larger inclination angle of $\alpha=60^{\circ}$ (Model C), the southeast equator directed more closely to the south, and the resulting southeast bump has a larger angular extension from the east to the southeastern edge in Model C compared with in Model A and B.  With $\alpha=40^{\circ}$ (Model A), three small protrusions appear in the northwest of the shell as a result of the RT fingers reaching the forward shock. Finally, in Model D, with a smaller spatial velocity of $v_{\mathrm s}=20 \, \rm km\,s^{-1}$, the wind bubble extends less deeply to the east, and, as a result, the southeast bump which locates more closely the simulation center is less prominent compared with the Model B. Moreover, a protrusion also arises around the north of the shell for the Model B and C as a result of the interaction of the ejecta with the wind bubble.

In the model, the location and the angular extension of the southeast bump, which is obvious in the detected X-ray/radio morphology of Tycho's SNR, depend heavily on the inclination angle, and it is constrained to  $\alpha\sim 50^{\circ}$ based on the simulations. With the wind velocity at the pole of $v_{\mathrm p} = 100\, \rm km\,s^{-1}$, the radial extension of the southeast bump on the resulting shell of the remnant determines that $v_{\mathrm s}$  should be larger than $20\, \rm km\,s^{-1}$, and a consistent extension of the bump can be reproduced with about $v_{\mathrm s}=0.3v_{\mathrm p}=30\, \rm km\,s^{-1}$.

\begin{figure}
\begin{center}
\includegraphics[width=0.5\textwidth]{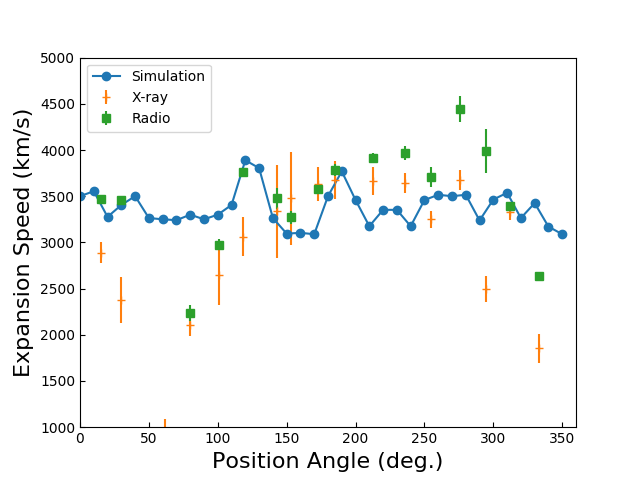}
\caption{The distribution of the speed of the forward shock on the position angle in the plane $x=0$ for the Model B at  $t = \mathrm{T_{age}} = 450\,\mathrm{yr}$. The detected velocities of different positions on the periphery of Tycho's SNR for a distance of $2.3\,\mathrm{kpc}$ based on the radio and the X-ray observations are also indicate for comparison \citep{Wea16}. \label{fig:speed}}
\end{center}
\end{figure}

Based on the long-term observations in the X-rays and the radio, the proper motion varies with locations around the periphery of Tycho's SNR, and a relation of the expansion velocities on the position angle $\theta$, which is the angle east of north with respect to geometric center, with a sinusoidal function is derived by \citet{Wea16}. Fig.\ref{fig:speed} shows the resulting expansion velocity of the forward shock on the position angle for the Model B at $t = \mathrm{T_{age}} = 450\,\mathrm{yr}$,  and the detected velocities of the proper motion from the X-rays and the radio in \citet{Wea16} for the distance of 2.3 kpc  are also indicated. The velocities from the simulation for Model B at different position angles of the forward shock range from $3000$ to $4000$\,$\mathrm{km}\,\mathrm{s}^{-1}$. For the shock velocities from the simulation and the derived ones from the X-ray/radio observation with the position angle, a significant  deviation exists between $50^{\circ}$  to $100^{\circ}$. Especially, the detected velocity at $\theta = 80^{\circ}$ is about $1000$\,$\mathrm{km}\,\mathrm{s}^{-1}$ smaller than the simulation. Moreover,  the velocities from $\theta = 200^{\circ}$  to $250^{\circ}$ are about $500$\,$\mathrm{km}\,\mathrm{s}^{-1}$ smaller than the detected ones. Aside from the variation, at $\theta = 15^{\circ}$  and $30^{\circ}$, the resulting velocities are well consistent with the the detected in the radio. From  $120^{\circ}$  to $150^{\circ}$, coincident with the protrusion to the southeast, the detected values of velocities and the decreasing trend are reproduced. The increasing trend, which is indicated in both the X-ray and the radio observations, is also consistent with the simulation of the velocities with the position angle from $150^{\circ}$  to $ 180^{\circ}$. Moreover,  except for $276^{\circ}$, the simulated velocities from  $250^{\circ}$  to $350^{\circ}$ are about $3500$\,$\mathrm{km}\,\mathrm{s}^{-1}$, which is in agreement with the averaged values of the detected ones in the radio and the X-ray band.

\section{Discussion and conclusions}
\label{DISCON}

A serials of 3D MHD simulations are performed in this paper both to investigate the formation mechanism of the peculiar morphology of Tycho's SNR and to derive the ambient material distribution before the supernova. These simulations are based on the assumption that the progenitor is imbedded in the bubble blown by the wind which has a latitude-dependent density distribution. Moreover, an inclination angle between the pole axis of the wind and the north direction, and a spatial motion of the progenitor are included to reproduce the asymmetry of the profile of the SNR shock.

With the pole velocity of $v_{\rm p}=100\rm \, km\,s^{-1}$, $\xi=20$ and the lasting time of $T_{\rm w}\sim10^5$~yr, the inclination angle of $\sim50^{\circ}$ between the pole of wind and north direction and the spatial velocity of $v_{\rm s}\sim0.3v_{\rm p}$ are derived based on the simulation to reproduce the detected profile of the outer border of the remnant. Furthermore, since a wind with a velocity of larger than $100 ~\mathrm{km\,s^{-1}}$ can be produced due to accretion of  onto a WD, and the SD scenario for the progenitor is compatible with main observed morphological feature under the assumptions used in this paper.

Based on the simulations with $v_{\rm p}=100\rm \, km\,s^{-1}$, $\dot{M}_{\rm w} = 10^{-7}\, {\rm M}_{\odot} {\rm yr}^{-1}$ and  $v_{\rm s}=0.3v_{\rm p}$, the wind bubble extends more deeply to the east with a longer evolution time, and the resulting southeast bump on the SNR shell distributes further away from the observed scale with an evolution time over $10^5$~yr for the wind bubble. The accretion onto the WD can last for a timescale large than $10^6$~yr to produce the large cavity with an extension of $13\mathrm{pc}\times27\mathrm{pc}$ around Tycho's SNR \citep{Cea16}. In this paper, we focus on the adjacent environment which can give effective influence on the morphology of the remnant, and the wind bubble, which is  driven by the wind during the time of  $\sim 10^5$~yr just before the supernova, has an extension of $\sim4\mathrm{pc}\times7\mathrm{pc}$. Therefore, the accretion wind may consists of two main phases. The former one produces the large cavity as indicated in radio observations, and the smaller cavity which determines the morphology of the remnant is formed in the later phase. Moreover, the model for the wind bubble in this paper is consistent with the common-envelope wind model proposed in \citet{MP17}.

\begin{figure}
\begin{center}
\includegraphics[width=0.5\textwidth]{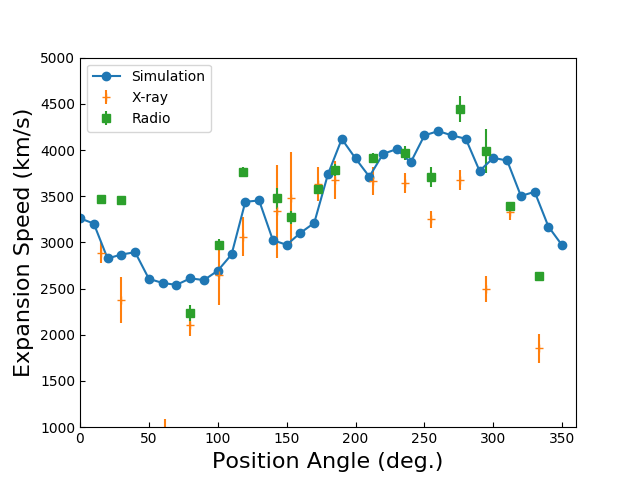}
\caption{The distribution of the speed of the forward shock on the position angle in the plane $x=0$ for the Model B at  $t = \mathrm{T_{age}} = 450\,\mathrm{yr}$ assuming the system has a velocity $v_{\rm k}=700\,\mathrm{km\,s}^{-1}$.  \label{fig:speedtwo}}
\end{center}
\end{figure}

The proper motion of the periphery of Tycho's SNR varies with the position angle, and the sinusoidal function, i.e., $F=A\sin(\theta+\phi)+Y$ with $A=0.062\pm0.0045\,\mathrm{arcsec}\,\mathrm{yr}^{-1}$ and $Y=0.293\pm0.002\,\mathrm{arcsec}\,\mathrm{yr}^{-1}$, had been obtained \citep{Wea16}. This trend is not reproduced by the model in this paper, and there is no significant departure from the geometric center to the explosion site. The peculiar trend can be naturally reproduced with the assumption that the ejecta has a spatial velocity of $v_{\rm sp}$ towards the southeast with respect to the Galaxy just after the supernova explosion, and the detected radial velocity of the blast wave is $v_{\rm r,bw} = v'_{\rm r,bw} - v_{\rm sp}\sin(\theta + \varphi)$, where $v'_{\rm r,bw}$ is the radial velocity of the blast wave with respect to the simulation center. The velocities of the forward shock in the simulation on different position angles vary around $v_{\rm ave}=3300\,\mathrm{km\,s}^{-1}$, and $v_{\rm sp}$ can be estimated to be $700\,\mathrm{km\,s}^{-1}$ based on $v_{\rm sp} = v_{\rm ave}A/Y$. Fig.\ref{fig:speedtwo} shows the resulting velocities of the forward shock on different position angles with $v_{\rm sp} = 700\,\mathrm{km\,s}^{-1}$ and $\varphi = 20^{\circ}$. Most of the detected velocities and the sinusoidal relation can be well reproduced. Moreover, the offset of the explosion site from the geometric center is $v_{\rm sp}\times T_{\mathrm{age}} = 0.32\,\mathrm{pc}$ if the spacial velocity $v_{\rm sp} = 700\,\mathrm{km\,s}^{-1}$ is in the real case.

\citet{WH09} investigated properties of the surviving companion star for the progenitor system composed of a white dwarf (WD) and a He star, and the companion star can obtain a high spatial velocity larger than $400\,\mathrm{km\,s}^{-1}$ because it has a relatively higher orbital velocity before the supernova compared with a main-sequence star. As a sequence, the WD can also have a spatial velocity with several hundreds of $\mathrm{km\,s}^{-1}$, and then this velocity is transferred to the ejecta after the explosion.

Based on observation with {\it Spitzer}, \citet{Wea13} found the density in the shell of Tycho's SNR had an order-of-magnitude variation over different position angles. In the simulation in this paper, the resulting densities in the shell varies insignificantly with the position angle, so more complicated ambient medium such as dense clumps should be included to reproduce the variation of the post-shock density around the periphery derived in \citet{Wea13}.

In this paper, we pay attention to the formation mechanism of the profile of the shock for Tycho's SNR, and the effects of the clumps in the ejecta and the shock acceleration process are not taken into account. The clumps with different velocities for Tycho's SNR have been clearly indicated based on the {\it Chandra}'s high angular resolution images \citep{SH17}.  \citet{Oea12} investigated the separation between the forward shock and the contact discontinuity of SN 1006 through 3D MHD simulation, and it was argued that the detected small separation and the protrusions for the remnant can be reproduced as a result of the clumping the ejecta.  Recently, the velocity of the ejecta knots of Tycho's SNR had been derived by \citet{Wea17} based on 3D measurements, and the velocities have a mean of 4430 km\,s$^{-1}$ with a spread from 2400 to 6600 km\,s$^{-1}$ for the assumed distance of $3.5$ kpc. Simulations had also been performed to discern whether the smooth or the clump ejecta profile can better explain the observed feature of the velocities of those knots, but the result showed that both assumptions can accommodate the 3D measurements \citep{Wea17}.

Alternatively, including the effect of the back-reaction of the accelerated charged particles, a smaller separation between the contact discontinuity and the forward shock can be also explained with a compression ratio larger than 4. Furthermore, there are evidences that the X-ray emissivity varies with different location of the forward shocks of both SN 1006 \citep{Rea13} and Tycho's SNR \citep{Lea15} due to the effect of shock obliquity. These effects can greatly influence the structure of the remnant. However, due to the complicated matter distribution resulting from the interacting between the wind and the ambient medium, protrusions can also be produced on the profile of the SNR shell as the Rayleigh-Taylor fingers approaching the forward shock.

\section*{Acknowledgements}
We thank the careful referee for the valuable comments and suggestions to greatly improve this paper.
We thank Dr. X.H. Sun for some discussion.
JF is partially supported by the
Natural Science Foundation of China (11563009),
the Yunnan Applied Basic Research
Projects (2016FB001), the Candidate Talents Training Fund of Yunnan Province (2017HB003)
and the Program for Excellent Young Talents, Yunnan University (WX069051). LZ acknowledges support from
NSFC (11433004). HY is partially supported by the Yunnan Applied
Basic Research Projects (2016FD105), the foundations of Yunnan Province (2016ZZX180, 2016DG006) and Kunming University (YJL15004, XJL15015).












\bsp	
\label{lastpage}
\end{document}